\documentclass[amsmath,amssymb,pra,aps,showpacs,superscriptaddress,twocolumn]{revtex4-1}
\usepackage{amsfonts,amsthm,graphics,bbm}
\usepackage[colorlinks=true,citecolor=blue,linkcolor=blue,urlcolor=blue]{hyperref}
\usepackage[usenames]{color}
\usepackage{graphicx}
\usepackage{subfigure}
\usepackage{amsmath}
\usepackage{epsfig}
\usepackage{dcolumn}
\usepackage{bm}
\usepackage{color}
\usepackage{times}
\usepackage{epstopdf}
\usepackage{amssymb}
\usepackage{amstext}
\usepackage{latexsym}
\usepackage{hyperref}
\usepackage{amsfonts}
\usepackage{psfrag}
\usepackage{txfonts}
\usepackage{float}
\usepackage{pifont}

\renewcommand{\Re}{\mathrm{Re}}

\newcommand{\ket}[1]{\vert #1 \rangle}

\newcommand{\braket}[2]{\langle #1 \vert #2 \rangle}

\begin{document}
	\preprint{APS/123-QED}
	\title{Swap Test with Quantum-Dot Charge Qubits}
	\author{Y.-D. Li}
	\affiliation{International Center of Quantum Artificial Intelligence for Science and Technology (QuArtist)\\  and Physics Department, Shanghai University, 200444 Shanghai, China}
	
	\author{N. Barraza}
	\affiliation{International Center of Quantum Artificial Intelligence for Science and Technology (QuArtist)\\  and Physics Department, Shanghai University, 200444 Shanghai, China}
	
	\author{G. Alvarado Barrios}
	\affiliation{International Center of Quantum Artificial Intelligence for Science and Technology (QuArtist)\\  and Physics Department, Shanghai University, 200444 Shanghai, China}
	
	\author{E. Solano}
\email[E. Solano]{\qquad enr.solano@gmail.com}
\affiliation{International Center of Quantum Artificial Intelligence for Science and Technology (QuArtist)\\  and Physics Department, Shanghai University, 200444 Shanghai, China}
\affiliation{IKERBASQUE, Basque Foundation for Science, Plaza Euskadi 5, 48009 Bilbao, Spain}
\affiliation{Kipu Quantum, Kurwenalstrasse 1, 80804 Munich, Germany}

\author{F. Albarr\'an-Arriagada}
\email[F. Albarr\'an-Arriagada]{\qquad pancho.albarran@gmail.com}
\affiliation{International Center of Quantum Artificial Intelligence for Science and Technology (QuArtist)\\  and Physics Department, Shanghai University, 200444 Shanghai, China}
	
\date{\today}

\begin{abstract}
	
We propose the implementation of the Swap Test using a charge qubit in a double quantum dot. The Swap Test is a fundamental quantum subroutine in quantum machine learning and other applications for estimating the fidelity of two unknown quantum states by measuring an auxiliary qubit. Our proposal uses a controlled three-qubit gate which is natural to quantum-dot charge qubits. It allows us to implement a Swap Test with a circuit depth of six layers, and an estimated gate time of less than 3 ns, that is below the coherence time of double quantum dots. This work paves the way for enhancing the toolbox of quantum machine learning developments in semiconductor qubits.

\end{abstract}

\maketitle
 
	\section{\label{sec:level1}Introduction}
Digital computation is one of the most outstanding scientific achievements of the 20th century, and it has dramatically changed technological production and daily life. The requirement of faster computation of increasingly complex problems have made the development of new paradigms for computing a priority for the scientific community. In this line, the emergence of quantum computing promises to surpass all the capabilities of classical computing, with recent claims of quantum advantage for random circuits~\cite{Arute2019Nature,Wu2021PRL} and boson sampling~\cite{Zhong2020Science,Zhong2021PRL}. However, the industrial application of quantum computing still appears far from the reach of current technology due to the small size of quantum processors and their noisy nature~\cite{Preskill2018Quantum}. 

During the last decades, different platforms have been proposed for the implementation of quantum computing protocols such as nuclear magnetic resonance~\cite{Kane1998Nature}, trapped ions~\cite{Cirac1995PRL,Sackett2001QIC}, optical setups~\cite{OBrien2007Science,Kok2007RMP}, quantum dots~\cite{Kloeffel2013ARCMP,Zhang2019NSR}, and superconducting circuits~\cite{Huang2020SCIS,Devoret2013Science}. The latter is a popular choice because of potential scalability, flexibility, and controllability. Nevertheless, quantum processors of superconducting circuits require cryogenic temperatures at around 20 mK, which poses a significant limitation for their industrial application for the near future. On the other hand, recent developments in fabrication techniques for quantum dots make them a promising candidate for future quantum processors. This is the case because such technology has considerably less stringent requirements as compared with superconducting circuits~\cite{Shi2013PRB,Dovzhenko2011PRB,Petersson2010PRL,Hayashi2003PRL,Shinkai2009PRL}. This has led to a growing interest from the community in the implementation of universal quantum gates in quantum dots based architectures. Quantum dots allows to build two types of qubits. The first one, known as semiconductor spin qubit, codifies the quantum bit in the spin degree of freedom of electrons trapped in a quantum dot~\cite{Loss1998PRA}. The second one, called charge qubit, codifies the qubit in the position of one electron trapped in a double quantum dot (DQD), which means that the state of the qubit is given by the label of the filled quantum dot~\cite{Gorman2005PRL,DiVincenzo2005Science}. Both architectures allow the implementation of a universal set of quantum gates and have good controllability.
	
Recently, different algorithms have been proposed for reaching quantum advantage in noisy intermediate-scale quantum (NISQ) devices for industrial applications, such as hybrid quantum variation algorithms~\cite{Cerezo2021NatRevPhys}, digitized adiabatic quantum computing~\cite{Hegade2021PRApplied}, and quantum machine learning protocols~\cite{Biamonte2017Nature}. For the last case, the most promising applications are related to big data algorithms such as classification or clustering problems, where the measurement of fidelity plays a fundamental role within the algorithms. One of the most popular algorithms for fidelity estimation between two quantum states by measuring an auxiliary qubit is the Swap Test, which is based on a three-qubit gate called Fredkin gate (controlled-swap). This gate has an optimal decomposition in one-qubit and two-qubit gates which involve the use of at least five two-qubit controlled gates~\cite{Smolin1996PRA}. Some platforms have allowed more efficient implementations of the Swap Test by the use of natural interactions, such as in optical setups~\cite{Milburn1989PRL,Kang2019SciRep}, or by making use of extra qubits, as in the case of trapped-ion implementations~\cite{Nguyen2021arXiv}.

In this work, we propose the implementation of the Swap Test in a system composed of three charge qubits by using a native controlled-three-qubit gate. This method allows us to implement the Swap Test with four three-qubit gates and four single-qubit rotations in less than $3$ ns, which is below the coherence time of double quantum dots. We test our protocol by calculating the fidelity of random states, which differ in phase, or differ in amplitude. All cases yield error in the order of $0.01$ on average, considering unitary evolutions. This work helps to the developing of machine learning algorithms using native gates in charge qubit quantum dots.
	
\section{\label{Sec02}The model}

\begin{figure}[t]
\centering
\includegraphics[width=0.95\linewidth]{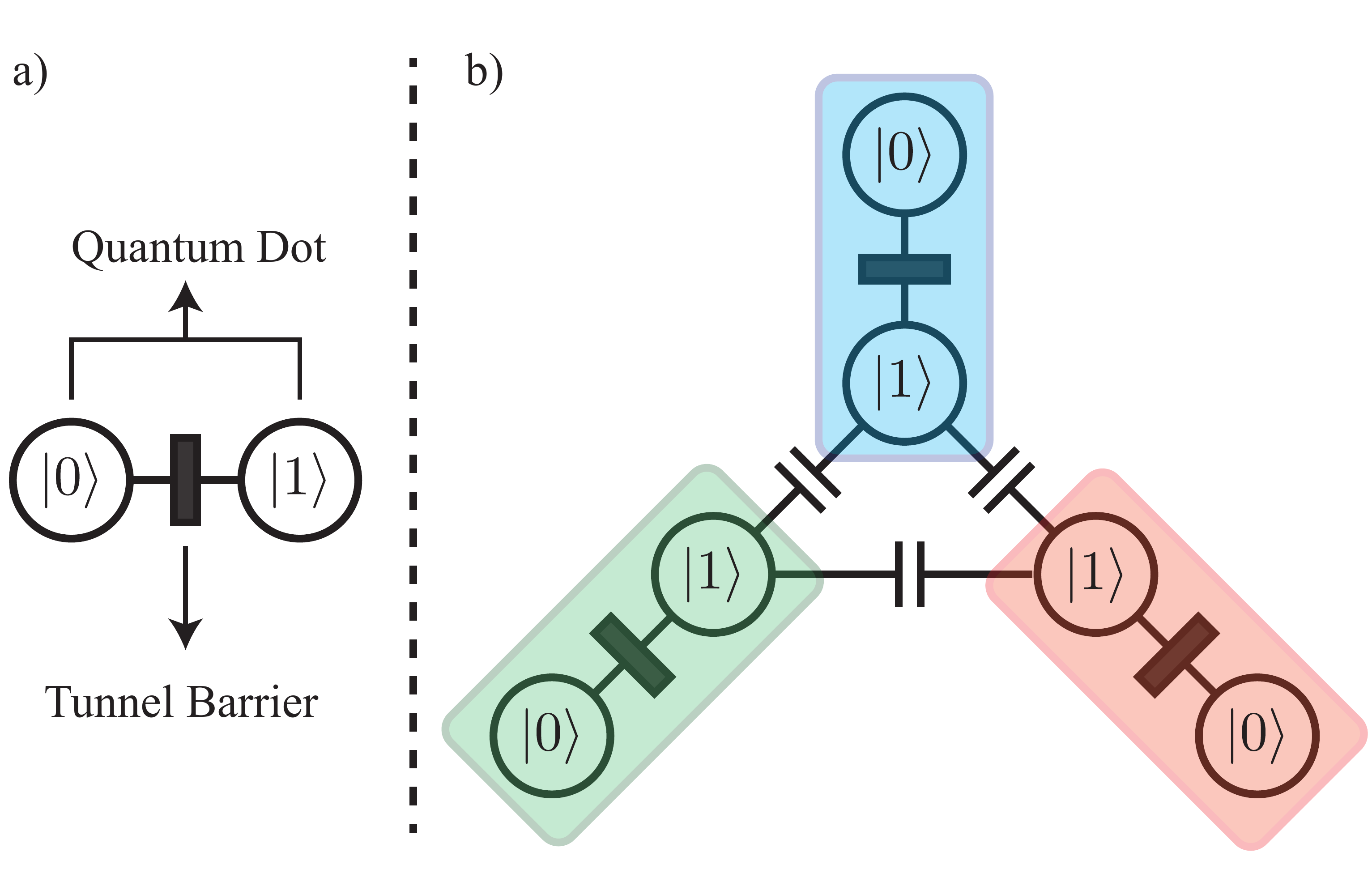}
\caption{a) Diagram of a quantum-dot charge qubit, it corresponds to a double-well potential where each minimum is given by a quantum dot separated by a tunnel junction. b) Diagram of the model described by the Hamiltonian of Eq.~(\ref{Eq01}). It corresponds to three quantum-dot charge qubits coupled through capacitances, where each qubit is highlighted by color boxes.}
\label{Fig01}
\end{figure}

\begin{figure}[b]
\centering
\includegraphics[width=0.8\linewidth]{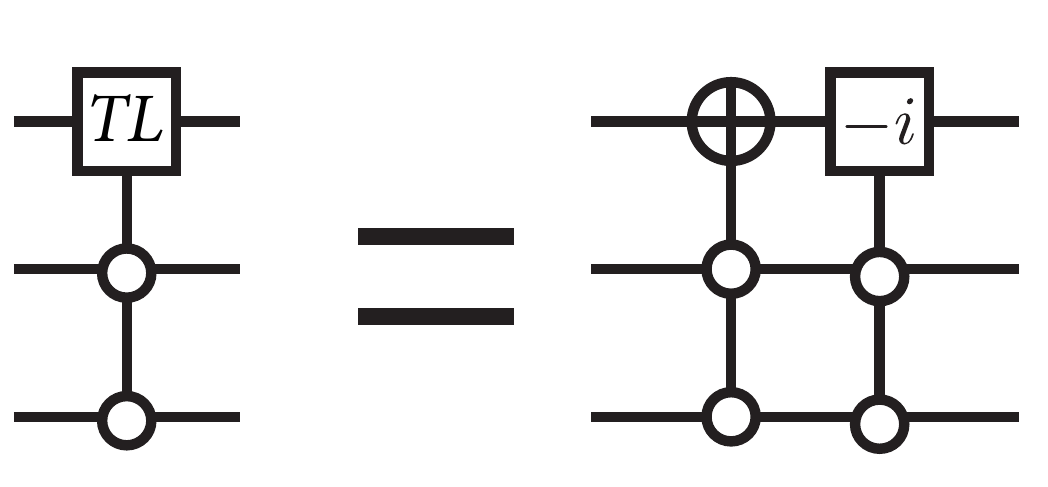}
\caption{Circuit representation of \textit{TL} gate and its relation with Toffoli gate and controlled-phase gate.}
\label{Fig02}
\end{figure}

Our system is composed of three charge-qubit quantum dots~\cite{Gorman2005PRL}, as in Fig.\ref{Fig01}a, coupled through capacitors as in Fig.~\ref{Fig01}b. A similar system has been experimentally implemented for a Toffoli-like (\textit{TL}) gate~\cite{Li2018PRApplied}, an essential resource in our algorithm to implement the Swap Test. The Hamiltonian of our system reads ($\hbar=1$)
\begin{eqnarray}
\label{Eq01}
H=&&\frac{\epsilon_1\sigma_{z,1}+\Delta_1\sigma_{x,1}}{2}+\frac{\epsilon_2\sigma_{z,2}+\Delta_2\sigma_{x,2}}{2}+\frac{\epsilon_3\sigma_{z,3}+\Delta_3\sigma_{x,3}}{2} \nonumber \\
&& + J_{12}\frac{I_1-\sigma_{z,1}}{2}\frac{I_2-\sigma_{z,2}}{2}+J_{13}\frac{I_1-\sigma_{z,1}}{2}\frac{I_3-\sigma_{z,3}}{2}\nonumber\\
&&+J_{23}\frac{I_2-\sigma_{z,2}}{2}\frac{I_3-\sigma_{z,3}}{2}.
\end{eqnarray}
where $\epsilon_j$ and $\Delta_j$ are the detuning energy and the inter-dot tunneling rate of the $j$th qubit, respectively. $J_{jk}$ is the inter-qubit coupling energy between the qubits $j$ and $k$. In addition, $\sigma_{\alpha,j}$ and $I_j$ are the Pauli-$\alpha$ matrix and identity operator for the $j$th qubit respectively. By controlling the external gate potential we can change the values of $\epsilon_j$, $\Delta_j$ and $J_{jk}$, which is fundamental for coherent manipulation with relatively high fidelity. Also, as shown in Fig. \ref{Fig01}, the qubits are encoded in the charge occupancies degree of freedom in each DQD, which can be measured by a quantum point contact. 

\section{\label{Sec03} Swap-Test algorithm}
\subsection{Three-qubit \textit{TL} gate}
First, we review the controlled-three-qubit gate reported in Ref.~\cite{Li2018PRApplied} that we call here \textit{TL} gate. This gate is a Toffoli-like gate, where the target qubit is flipped if the control qubits are in $\ket{0}$ and a factor $-i$ is added, all other control combinations do not change the target qubit state. Table~\ref{Tab01} summarizes the action of this gate with the first qubit as target, and the other two as control, and Fig.~\ref{Fig02} shows the circuit representation and relation with the Toffoli gate.

\begin{table}[b] 
\begin{center}
\begin{tabular}{|c|c|} 
\hline
input state&output state\\ [0.5ex]
\hline
$\ket{000}$&$-i\ket{100}$\\
$\ket{001}$&$\ket{001}$\\
$\ket{010}$&$\ket{010}$\\
$\ket{011}$&$\ket{011}$\\
$\ket{100}$&$-i\ket{000}$\\
$\ket{101}$&$\ket{101}$\\
$\ket{110}$&$\ket{110}$\\
$\ket{111}$&$\ket{111}$\\
\hline
\end{tabular}
\end{center}
\caption{\textit{TL}-gate truth table}
\label{Tab01}
\end{table}

\begin{figure*}[t]
\includegraphics[width=1.03\linewidth]{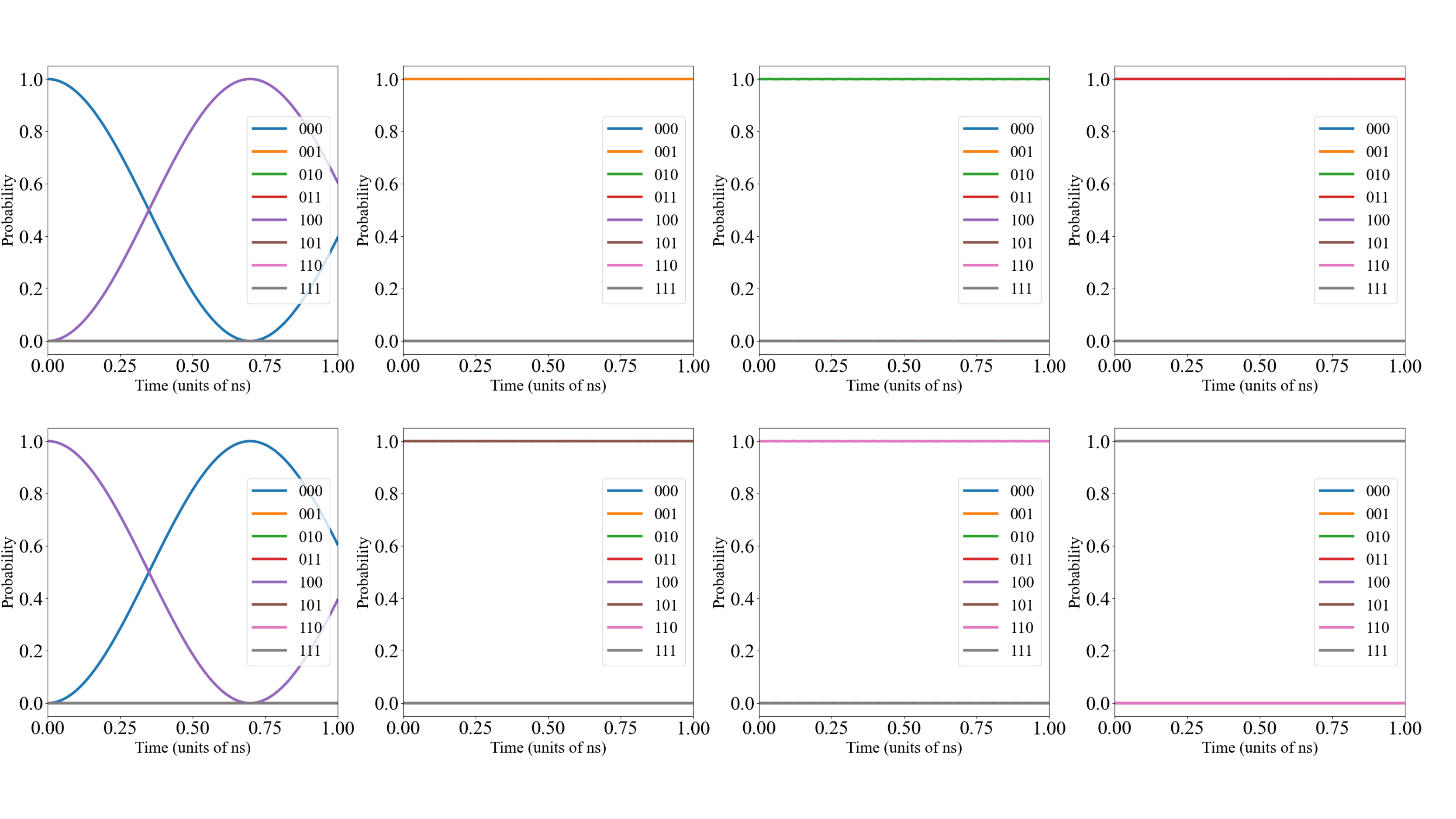}	
\caption{Time evolution given by Hamiltonian~(\ref{Eq01}) to perform \textit{TL} gate. The input states are (from left to right, from up to down): $|000\rangle$, $|001\rangle$, $|010\rangle$, $|011\rangle$, $|100\rangle$, $|101\rangle$, $|110\rangle$ and $|111\rangle$.}
\label{Fig03}
\end{figure*}

To perform this gate, we make use of the Coulomb blockade produced by the state $\ket{1}$ in the control qubit, which implies that extra energy is required to flip a qubit capacitively coupled to the control. We consider qubit $1$ as target qubit, the label of the qubits is arbitrary. Now, we can manipulate the external voltages to obtain the regime $|\Delta_1| {\gg} |\epsilon_1|$, $|\epsilon_2| {\gg} |\Delta_2|$, $|\epsilon_3| {\gg} |\Delta_3|$, and $\epsilon_2\approx\epsilon_3\gg J_{23}$, where the last condition implies that we can neglect the interaction between the qubits $2$ and $3$. In this regime, the Hamiltonian of Eq. (\ref{Eq01}) reduces to 
\begin{eqnarray}
\label{Eq02}
H=&&\frac{\Delta_1\sigma_{x,1}}{2}+\frac{\epsilon_2\sigma_{z,2}}{2}+\frac{\epsilon_3\sigma_{z,3}}{2}+J_{12}\frac{I_1-\sigma_{z,1}}{2}\frac{I_2-\sigma_{z,2}}{2}\nonumber\\
&&+J_{13}\frac{I_1-\sigma_{z,1}}{2}\frac{I_3-\sigma_{z,3}}{2}.
\end{eqnarray}

Since in Eq.~(\ref{Eq02}), the operators for qubit $2$ and qubit $3$ are only $\sigma_z$, we can write the eigenvectors of this Hamiltonian as $\ket{\psi_{\pm}}\ket{jk}$, where $\ket{\psi_{\pm}}$ are given by
\begin{eqnarray}
\label{Eq03}
\!\!\!\!\!\! \ket{\psi_\pm}=\sqrt{\frac{4(\lambda_{\pm}+m)^2}{4(\lambda_{\pm}+m)^2+\Delta_1^2}}\ket{1}+\sqrt{\frac{\Delta_1^2}{4(\lambda_{\pm}+m)^2+\Delta_1^2}}\ket{0},
\end{eqnarray}
with
\begin{eqnarray}
\label{Eq04}
&&m=-\frac{\epsilon_2}{2}(-1)^j-\frac{\epsilon_3}{2}(-1)^k,\quad n=-J_{12}\delta_{1j}-J_{13}\delta_{1k},
\end{eqnarray}
and 
\begin{eqnarray}
\label{Eq05}
\lambda_{\pm}=\frac{-\left(2m+n\right)\pm\sqrt{n^2+\Delta_1^2}}{2}
\end{eqnarray}	
are the corresponding eigenvalues. Using the spectral decomposition, it is easy to calculate the evolution of the states in the computational basis, obtaining
\begin{eqnarray}
\label{Eq06}
e^{-iHt}\ket{1jk}=&&e^{-i\xi t} \Bigg\{ \left[\cos\left(\omega t\right)+\frac{n}{\sqrt{n^2+\Delta_1^2}}i\sin\left(\omega t\right)\right]\ket{1jk} \nonumber\\
&&-\frac{\Delta_1}{\sqrt{n^2+\Delta_1^2}}i\sin\left(\omega t\right)\ket{0jk} \Bigg\} , \nonumber\\
e^{-iHt}\ket{0jk}=&&e^{-i\xi t} \Bigg\{ \frac{-\Delta_1}{\sqrt{n^2+\Delta_1^2}}i{\sin}(\omega t)\ket{1jk}  \nonumber\\
&&+\Bigg[{\cos}\left(\omega t\right)-\frac{n}{\sqrt{n^2+\Delta_1^2}}i{\sin}\left(\omega t\right)\Bigg]\ket{0jk}  \Bigg\},
\end{eqnarray}
with 
\begin{equation}
\label{Eq07}
\xi=\frac{-\left(2m+n\right)}{2},\,\omega=\frac{\sqrt{n^2+\Delta_1^2}}{2} \, .
\end{equation}
Now, if $J_{12}\sim J_{13}\gg\Delta_1$, we obtain
\begin{eqnarray}
\label{Eq08}
e^{-iHt}\ket{1jk}=&&e^{im t} \Bigg\{ \left[\cos\left(\omega t\right)\right]\ket{1jk}-i\sin\left(\omega t\right)\ket{0jk} \Bigg\},\nonumber\\
e^{-iHt}\ket{0jk}=&&e^{im t} \Bigg\{-i{\sin}(\omega t)\ket{1jk}+{\cos}\left(\omega t\right)\ket{0jk}  \Bigg\},
\end{eqnarray}
for $j=k=0$, and 
\begin{eqnarray}
\label{Eq09}
e^{-iHt}\ket{1jk}=&&e^{-i(\xi-\omega) t}\ket{1jk}=e^{i(m+n) t}\ket{1jk} ,\nonumber\\
e^{-iHt}\ket{0jk}=&&e^{-i(\xi+\omega) t}\ket{0jk}=e^{im t}\ket{0jk}
\end{eqnarray}
for other cases. We note that the global phase in each transformation can be deleted in the interaction picture by choosing the free Hamiltonian 
\begin{eqnarray}
\label{Eq10}
H_0=&&\frac{\epsilon_2\sigma_{z,2}}{2}+\frac{\epsilon_3\sigma_{z,3}}{2}+J_{12}\frac{I_1-\sigma_{z,1}}{2}\frac{I_2-\sigma_{z,2}}{2}\nonumber\\
&&+J_{13}\frac{I_1-\sigma_{z,1}}{2}\frac{I_3-\sigma_{z,3}}{2} \, .
\end{eqnarray}
Therefore, in the interaction picture, after a time $t_T=\frac{\pi}{\Delta_1}$, we obtain the desired \textit{TL}-gate transformation summarized in table~\ref{Tab01}.

Experimentally, the parameters can be fixed as $\epsilon_1=0$ GHz, $\epsilon_2=\epsilon_3=-303.854$ GHz, $\Delta_1=\Delta_2=4.5$ GHz, $\Delta_3=1$ GHz, $J_{12}=159.523$ GHz,  $J_{13}=205.101$ GHz and $J_{23}=0$ GHz. This yields a gate time $t_T\sim0.7$ ns. Figure~\ref{Fig03} shows the probability evolution of each state in the computational basis, with the initial state being the eight different states of the form $\ket{jkl}$ with $j,k,l\in \{0,1\}$ and with the Hamiltonian of Eq.~(\ref{Eq10}). As the inter-qubit coupling can be tuned, we can perform the \textit{TL} gate for any qubit as target~\cite{Yu2016Nano}.			
	
\subsection{$S$ gate}
\begin{figure}[b]
\centering
\includegraphics[width=1\linewidth]{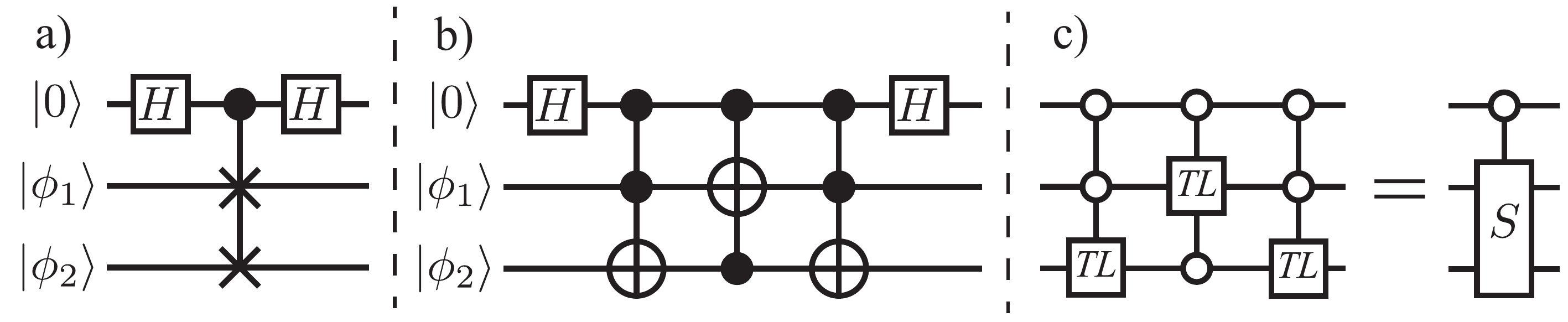}
\caption{a) Circuit algorithm for Swap Test using a controlled-swap gate. b) circuit diagram for Swap Test using Toffolli gates. c) $S$ gate and its decomposition in \textit{TL} gates.}
\label{Fig04}
\end{figure}
\begin{figure*}[t]		
\centering										
\includegraphics[width=1.02\linewidth]{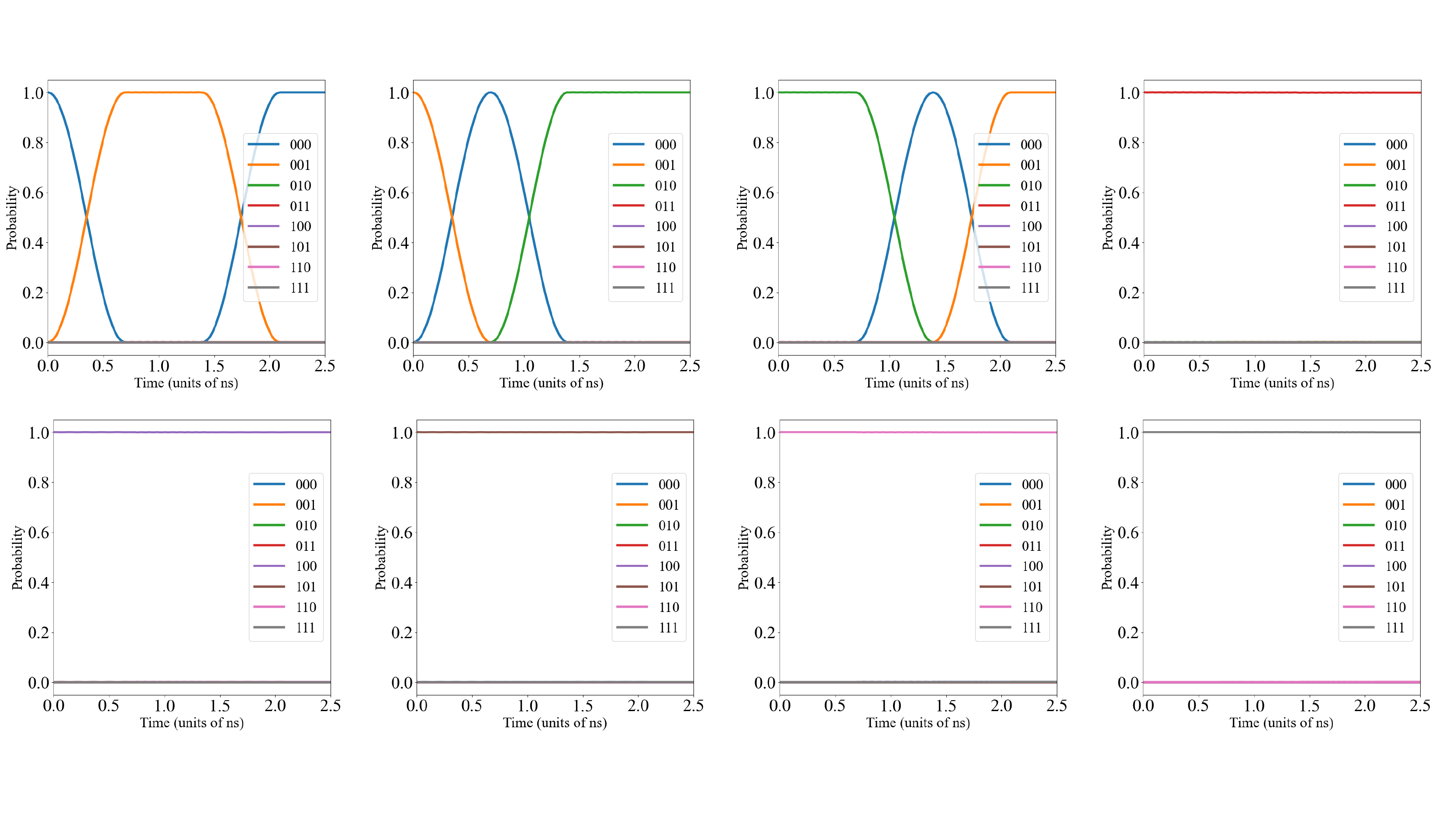}
\caption{Time evolution to perform $S$ gate composed of three \textit{TL} gates exchanging the control qubits. The input states are (from left to right, from up to down): $|000\rangle$, $|001\rangle$, $|010\rangle$, $|011\rangle$, $|100\rangle$, $|101\rangle$, $|110\rangle$ and $|111\rangle$.}
\label{Fig05}
\end{figure*}

It is well known that to perform a Swap Test, a key element is the controlled-swap or Fredkin gate, which is obtained with three Toffoli gates (see Fig.~\ref{Fig04} a and b). Now, since our \textit{TL} gate is an imperfect Toffoli gate, by merging three of them we can obtain an imperfect controlled-swap gate that we call $S$ gate. Figure~\ref{Fig04} c shows the circuit of the $S$ gate in terms of the \textit{TL} gate. The action of this gate under the eight computational basis states is summarized in Table~\ref{Tab02}. We can see it is close to a controlled-swap gate but with a factor $-1$ in the states $\ket{000}$, $\ket{010}$ and $\ket{001}$, where the first qubit is the control qubit. As this gate is composed of three \textit{TL} gates, the gate time is $t_S=3t_T\sim 2.1$ ns. Figure~\ref{Fig05} shows the probability of each $\ket{jkl}$ state using the computational basis as initial states. Here, we can see the population inversion that defines a controlled-swap-like gate.  
			
\begin{table}[b] 
\begin{center}
\begin{tabular}{|c|c|} 
\hline
input state&output state\\ [0.5ex]
\hline
$\ket{000}$&$-\ket{000}$\\
$\ket{001}$&$-\ket{010}$\\
$\ket{010}$&$-\ket{001}$\\
$\ket{011}$&$\ket{011}$\\
$\ket{100}$&$\ket{000}$\\
$\ket{101}$&$\ket{101}$\\
$\ket{110}$&$\ket{110}$\\
$\ket{111}$&$\ket{111}$\\
\hline
\end{tabular}
\end{center}
\caption{$S$-gate truth table}
\label{Tab02}
\end{table}

\subsection{Swap Test}
\begin{figure}[b]
\centering
\includegraphics[width=0.9\linewidth]{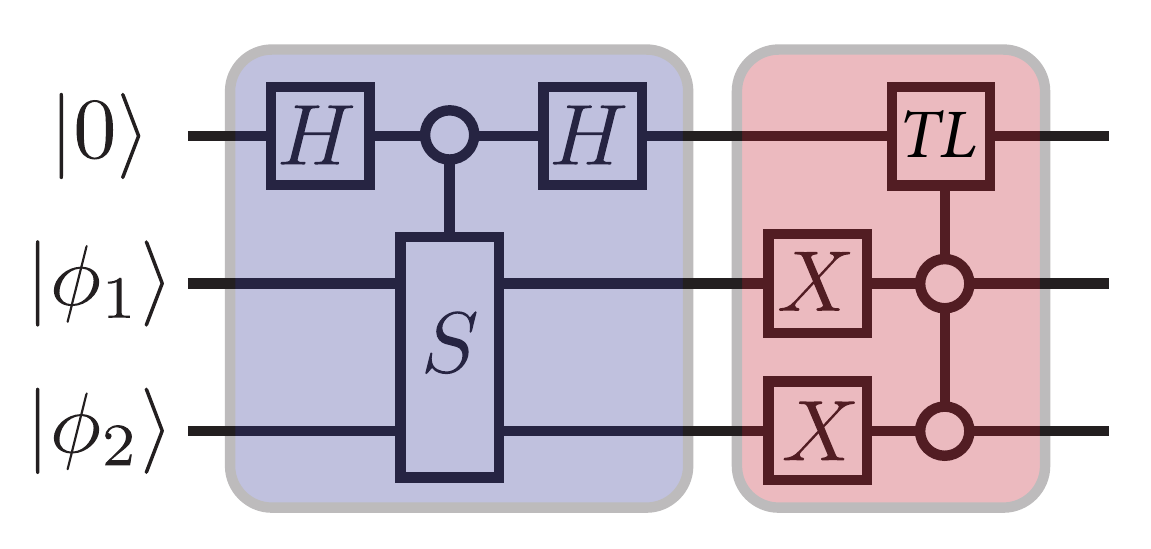}
\caption{Circuit diagram for Swap Test.}
\label{Fig06}
\end{figure}

We will separate our algorithm in two parts to perform the Swap Test. First, we perform the same algorithm as in Fig.~\ref{Fig04}a. but we replace the controlled-swap gate with the $S$ gate. This is shown in Fig.~\ref{Fig06} and highlighted with blue. As the $S$ gate is an imperfect controlled-swap gate, after this step we obtain an imperfect Swap Test, then the idea of the second part of the algorithm, highlighted with red in Fig.~\ref{Fig06}, is to correct the occupancy probabilities of the auxiliary qubit, in order to recover the result of the Swap Test.

To understand how the algorithm works, consider the initial three-qubit state $\ket{\psi_0}=\ket{0}\ket{\phi_1}\ket{\phi_2}$, where
\begin{equation}
\ket{\phi_j}=\alpha_j\ket{0}+\beta_j\ket{1} .
\end{equation}
Then, we obtain
\begin{equation}
\ket{\psi_0}=\ket{0}\left(\alpha_1\alpha_2\ket{00}+\alpha_1\beta_2\ket{01}+\beta_1\alpha_2\ket{10}+\beta_1\beta_2\ket{11}\right).
\end{equation}

After the first part of the algorithm (blue box) we obtain the state $\ket{\psi_1}$ given by
\begin{eqnarray}
\ket{\psi_1}=&&\frac{1}{2}\bigg\{\ket{0}[(\alpha_1\beta_2-\beta_1\alpha_2)\ket{01}+(\beta_1\alpha_2-\alpha_1\beta_2)\ket{10}\nonumber\\
&&+2\beta_1\beta_2\ket{11}]+\ket{1}[-(\alpha_1\beta_2+\beta_1\alpha_2)\ket{01}\nonumber\\
&&-(\beta_1\alpha_2+\alpha_1\beta_2)\ket{10}-2\alpha_1\alpha_2\ket{00}]\bigg\}.
\end{eqnarray}
At this stage, the probabilities of obtaining $\ket{0}$ and $\ket{1}$, which we denote by $P_0$ and $P_1$, respectively, can be written as
\begin{eqnarray}
P_0=&&\frac{1}{4}\Big[4|\beta_1\beta_2|^2+2|\alpha_1\beta_2|^2+2|\beta_1\alpha_2|^2\nonumber\\
&&-4\Re(\alpha_1\alpha_2\beta_1\beta_2)\Big]\nonumber\\
P_1=&&\frac{1}{4}\Big[4|\alpha_1\alpha_2|^2+2|\alpha_1\beta_2|^2+2|\beta_1\alpha_2|^2\nonumber\\
&&+4\Re(\alpha_1\alpha_2\beta_1\beta_2)\Big] \, .
\end{eqnarray}
Now, subtracting there probabilities we obtain
\begin{equation}
P_1-P_0=|\alpha_1\alpha_2|^2+2\Re(\alpha_1\alpha_2\beta_1\beta_2)-|\beta_1\beta_2|^2 .
\end{equation}
This expression differs from the fidelity between $\ket{\phi_1}$ and $\ket{\phi_2}$ in the sign of the last term. To fix this issue, we need that the term $|\beta_1\beta_2|^2$ appears in the probability $P_1$. This is achieved with the second part of the algorithm, where we flip the first qubit when the other two are in the state $\ket{11}$. This means that first we flip the last two qubits and then we perform the \textit{TL} gate with the first qubit as the target. After applying the second part of the algorithm we obtain the state $\ket{\psi_2}$ given by
\begin{eqnarray}
\ket{\psi_2}=&&\frac{1}{2}\bigg\{\ket{0}[(\alpha_1\beta_2-\beta_1\alpha_2)\ket{10}+(\beta_1\alpha_2-\alpha_1\beta_2)\ket{01}]\nonumber\\
&&-\ket{1}[2i\beta_1\beta_2\ket{00}+(\alpha_1\beta_2+\beta_1\alpha_2)\ket{10}\nonumber\\
&&+(\beta_1\alpha_2+\alpha_1\beta_2)\ket{01}+2\alpha_1\alpha_2\ket{11}]\bigg\},
\end{eqnarray}
where the probabilities are given by
\begin{eqnarray}
P_0=&&\frac{1}{4}\Big[2|\alpha_1\beta_2|^2+2|\beta_1\alpha_2|^2-4\Re(\alpha_1\alpha_2\beta_1\beta_2)\Big]\nonumber\\
P_1=&&\frac{1}{4}\Big[4|\alpha_1\alpha_2|^2+2|\alpha_1\beta_2|^2+2|\beta_1\alpha_2|^2\nonumber\\
&&+4\Re(\alpha_1\alpha_2\beta_1\beta_2)+4|\beta_1\beta_2|^2\Big] .
\end{eqnarray}
Now, subtracting these two quantities we obtain
\begin{equation}
P_1-P_0=|\alpha_1\alpha_2|^2+2\Re(\alpha_1\alpha_2\beta_1\beta_2)+|\beta_1\beta_2|^2=|\braket{\phi_1}{\phi_2}|^2 .
\end{equation}
This means that we can correctly estimate the fidelity between two unknown states by measuring the probabilities of the auxiliary qubit state, $P_0$ and $P_1$. 

Notice that we use Hadamard and $X$ gates (single-qubit gates) in our protocol. Current technology in double quantum dots allows us to control a single qubit by pulses in the gate voltage, performing universal single-qubit gates in the order of tens of ps, which translates to very high fidelity for current devices. This means that we can use perfect Hadamard gates and $X$-gates to test our protocol without too much error~\cite{Cao2013NatCommun}.

\begin{figure}[t]
\centering
\includegraphics[width=1\linewidth]{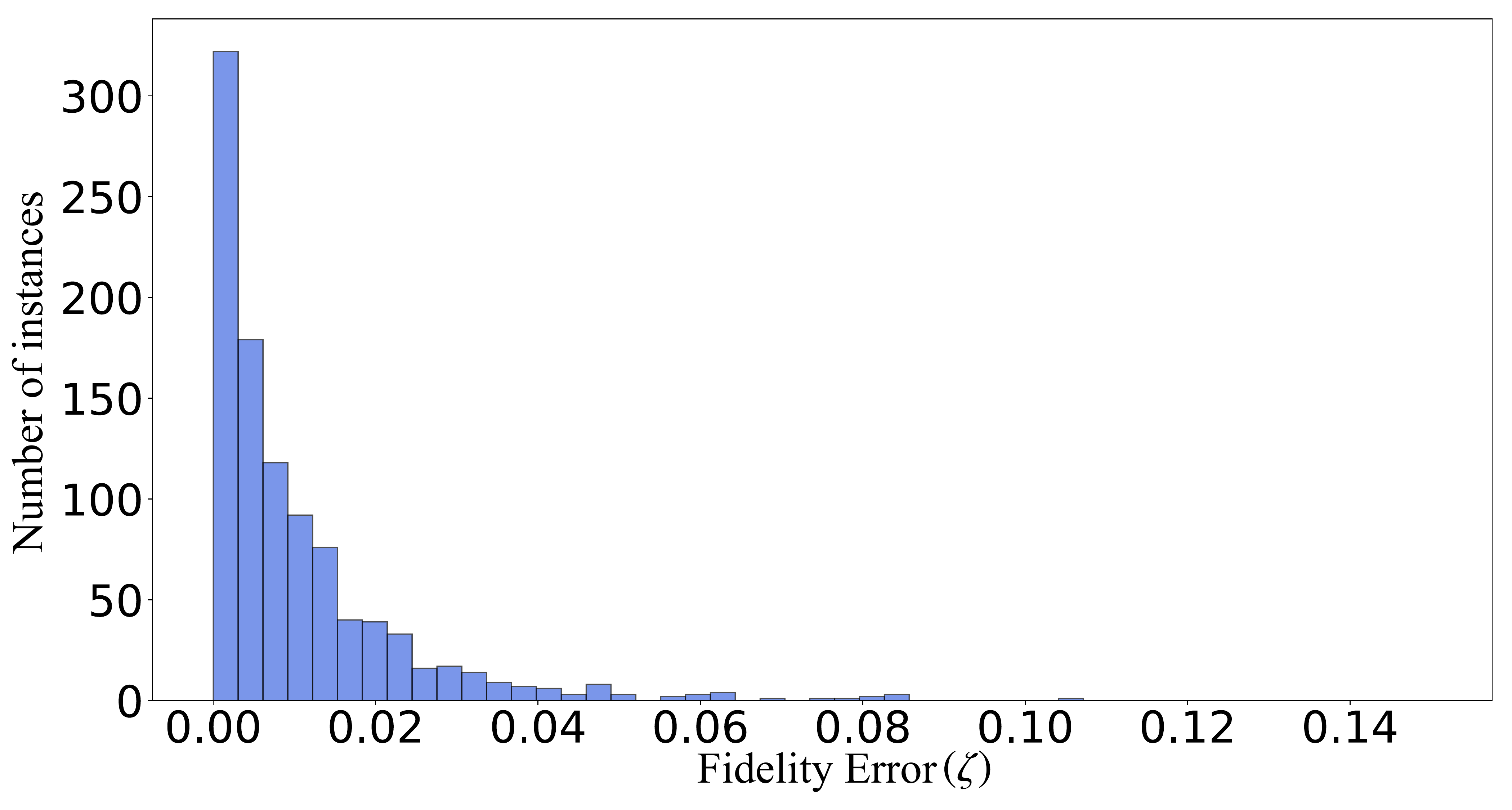}
\caption{Hystogram for fidelity error ($\zeta$) defined in Eq.~(\ref{Eq19}) for the case of 1000 random instances.}
\label{Fig07}
\end{figure}

\begin{figure}[t]
\centering
\includegraphics[width=1\linewidth]{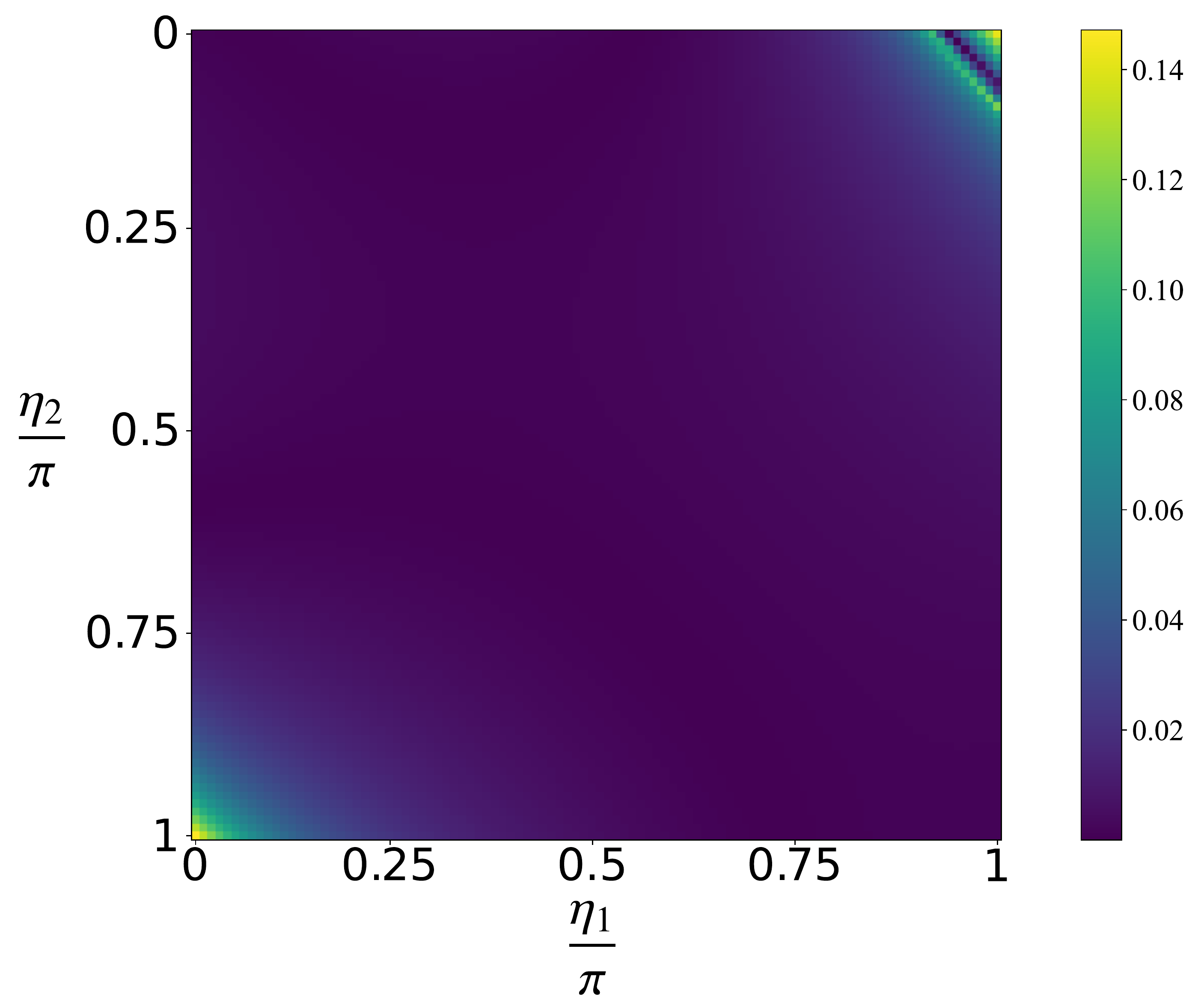}
\caption{Color map plot for fidelity error ($\zeta$) defined in Eq.~(\ref{Eq19}) for the phase difference case.}
\label{Fig08}
\end{figure}
 
To test our algorithm, our figure of merit is given by the fidelity error
\begin{equation}
\label{Eq19}
\zeta=|(P_0-P_1)-|\braket{\phi_1}{\phi_2}|^2| \, ,
\end{equation}
where $P_0$ and $P_1$ are calculated from our algorithm. Here, the \textit{TL} gates are given by the time evolution generated by Hamiltonian in Eq.~(\ref{Eq01}), using the corresponding parameters for target and control qubits. The single-qubit gates are assumed to be perfect gates, which is close to the current status of the experiments. We probe our Swap Test for three different cases:
\begin{enumerate}
\item Random states: we consider 1000 random instances, each instance involves a random state for $\ket{\phi_1}$ and another random state for $\ket{\phi_2}$. The results of the fidelity error are collected in the histogram of Fig.~\ref{Fig07}. From the figure, we can see that in most cases the error is below $0.02$, with almost no instances beyond $\zeta=0.06$.

\item Phase difference: we consider input states given by
\begin{equation}
\ket{\phi_j}=\sqrt{\frac{1}{2}}\left(\ket{0}+e^{i\eta_j}\ket{1}\right) ,
\end{equation}
where $\eta_j\in\{0,\pi\}$. Figure~\ref{Fig08} shows the color map of the fidelity error $\zeta$ as a function of the angles $\eta_1$ and $\eta_2$. We can see that the error is very small in the vast majority of cases, except in the corners where the the initial states are orthogonal, and the error approaches 0.14.

\begin{figure}[t!]
\centering
\includegraphics[width=1\linewidth]{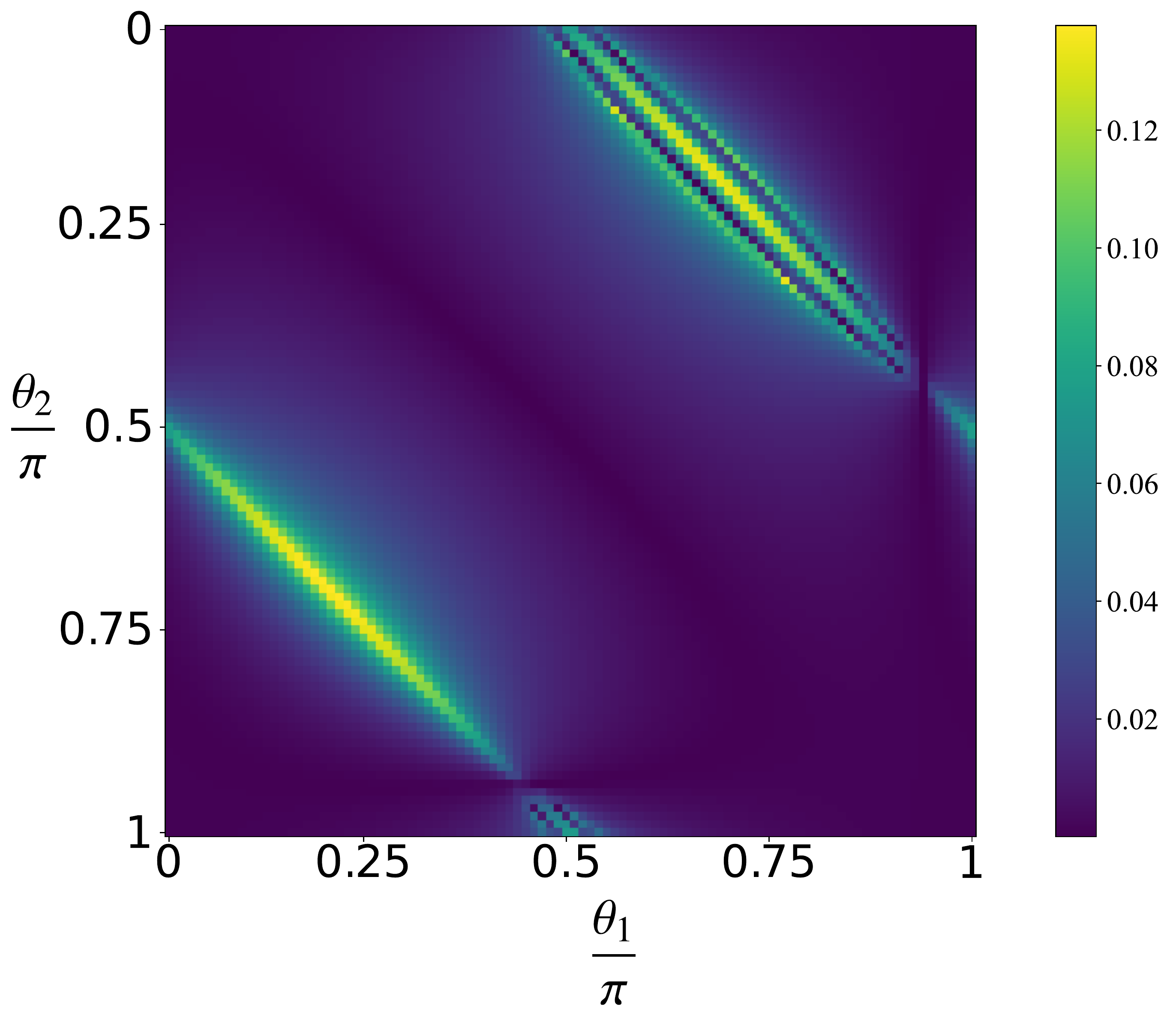}
\caption{Color map plot for fidelity error ($\zeta$) define in Eq.~(\ref{Eq19}) for the amplitude difference case.}
\label{Fig09}
\end{figure}

\item Amplitude difference: we consider input states given by
\begin{equation}
\ket{\phi_j}=\cos(\theta_j)\ket{0}+\sin(\theta_j)\ket{1} ,
\end{equation}
where $\theta_j\in\{0,\pi\}$. Figure~\ref{Fig09} shows the color map of the fidelity error as a function of the angles $\theta_1$ and $\theta_2$. Similarly to the previous case, we can see that the error is very small for most cases, and the regions of higher error ($\zeta\sim 0.13$) correspond to the configurations where the input states are orthogonal.
\end{enumerate}

Therefore, the proposed algorithm for the Swap Test can estimate the fidelity with an average error of $0.01$, becoming less accurate for orthogonal states. Finally, we mention that the duration of our algorithm is essentially the time required for the four three-qubit gates, around $t\approx2.8$ ns. Thus, our algorithm can be performed within the coherence time reported by the community for this class of qubits, which can reach $10$ ns~\cite{Petersson2010PRL}. Therefore, our protocol is experimentally feasible and could be useful to promote this architecture for the efficient implementation of quantum machine learning protocols.

\section{\label{Sec04}Conclusion}
We have proposed  the implementation of a Swap Test with charge qubits in double quantum dots. We use an experimentally-demonstrated controlled three-qubit gate (\textit{TL} gate), similar to the Toffoli gate. This allows us to implement an imperfect controlled-swap gate that we termed $S$ gate, which is at the core of the proposed protocol, performing the Swap Test using natural gates for double quantum dots. Our protocol is able to estimate the fidelity between two unknow qubit states with an average error of $0.01$ for a unitary evolution. The estimated duration is of less than $3$ ns, which is bellow the coherence time that can be achieved with current devices. This proposal for a Swap Test is a step forward in the efforts for semiconductors qubits to efficiently perform quantum machine learning and quantum computing algorithms.

\section{Acknowledgments}

The authors acknowledge the financial support from projects STCSM (2019SHZDZX01-ZX04 and 20DZ2290900), and SMAMR (2021-40).

\end{document}